# Comparative Analysis of OECD Countries Based on Energy Trilemma Index: A Clustering Approach

**Emre AKUSTA** [1]

**Abstract**

*This study analyzes OECD countries in the context of the energy trilemma index and clusters countries with similar characteristics. In the study, the k-means clustering technique is used. The optimum number of clusters was determined using the Elbow method in combination with the Silhouette Index. Moreover, all results are visualized to enhance comprehensibility. The results show that countries such as Austria, Canada, Finland, and Denmark are in the high energy trilemma group with index scores of 82.2, 82.3, 82.7, and 83.3, respectively. Countries in the high group have achieved a high level of balance between energy security, energy equity, and environmental sustainability. In addition, countries such as Belgium, Hungary, Australia, the Czech Republic, and Estonia are in the medium energy trilemma group with index scores of 76.4, 76.6, 77.1, 77.6, and 78.7, respectively. Countries in the medium group have made progress in balancing the dimensions of the energy trilemma but have not yet reached excellence. However, countries such as Mexico, Türkiye, Colombia, and Costa Rica are in the low energy trilemma group with index scores of 63.1, 64.1, 64.8, and 69.3, respectively. These low energy trilemma group countries face significant challenges in balancing energy security, energy equity, and environmental sustainability and need to make improvements in these areas.*
***Keywords:*** *Energy Trilemma, OECD, K-means, Cluster Analysis, Elbow Method*
***Jel Codes:*** *Q30, Q40, Q50*

## *OECD Ülkelerinin Enerji Trilemma Endeksine Dayalı Karşılaştırmalı Analizi: Kümeleme Yaklaşımı*

***Özet***

*Bu çalışma, OECD ülkelerini enerji trilemma endeksi bağlamında analiz etmekte ve benzer özelliklere sahip ülkeleri kümelemektedir. Çalışmada k-ortalamalar kümeleme tekniği kullanılmıştır. Optimum küme sayısı, Siluet Endeksi ile birlikte Dirsek yöntemi kullanılarak belirlenmiştir. Ayrıca, anlaşılabilirliği artırmak için tüm sonuçlar görselleştirilmiştir. Sonuçlar Avusturya, Kanada, Finlandiya ve Danimarka gibi ülkelerin sırasıyla 82.2, 82.3, 82.7 ve 83.3 endeks puanlarıyla yüksek enerji trilemma grubunda yer aldığını göstermektedir. Yüksek grupta yer alan ülkeler enerji güvenliği, enerji eşitliği ve çevresel sürdürülebilirlik arasında yüksek düzeyde denge sağlamışlardır. Ayrıca Belçika, Macaristan, Avustralya, Çek Cumhuriyeti ve Estonya gibi ülkeler sırasıyla 76.4, 76.6, 77.1, 77.6 ve 78.7 endeks puanlarıyla orta enerji trilemma grubunda yer almaktadır. Orta gruptaki ülkeler enerji trilemmasının boyutlarını dengeleme konusunda ilerleme kaydetmiş ancak henüz mükemmelliğe ulaşamamıştır. Ancak Meksika, Türkiye, Kolombiya ve Kosta Rika gibi ülkeler sırasıyla 63,1, 64,1, 64,8 ve 69,3 endeks puanlarıyla düşük enerji trilemma grubunda yer almaktadır. Bu düşük enerji trilemma grubu ülkeleri enerji güvenliği, enerji eşitliği ve çevresel sürdürülebilirliği dengelemede önemli zorluklarla karşı karşıyadır ve bu alanlarda iyileştirmeler yapmaları gerekmektedir.*

---

[1] Assist. Prof., Kırklareli University, Faculty of Economics and Administrative Sciences, Department of Economics, Kirklareli, Türkiye **EMAIL:** emre.akusta@klu.edu.tr **ORCID:** 0000-0002-6147-5443

## 1. INTRODUCTION

It is generally accepted that energy is a critical input in key sectors such as industry, housing, agriculture, and transportation. More efficient use of energy by these sectors provides significant environmental benefits by reducing environmental risks. In particular, the use of fossil fuels leads to environmental problems such as greenhouse gas emissions and global warming. In this regard, private and public institutions and organizations emphasize the importance of reducing the harmful environmental impacts of energy consumption (Suranovic, 2013).

Climate change leads to an increase in global average temperatures through weather and ocean cycles, causing various natural phenomena to occur and human activities to contribute to global warming. The increase in greenhouse gases is the main cause of these interactions. As the world economy develops, the economic activities and positive production of all countries are closely linked, leading to significant changes in energy consumption and thus greenhouse gas emissions (Sueyoshi & Goto, 2013). Furthermore, energy demand, driven by population growth, urbanization, and industrialization, is expected to increase by one-third between 2015 and 2040 (Lee et al., 2018). This increase highlights the challenge of balancing energy security and environmental sustainability, especially for fossil fuel-dependent countries (Iqbal et al., 2019; Mohsin et al., 2019a).

The energy sector needs to transform from a fossil fuel-based system to a zero-carbon emission sector (IRENA, 2018). This transformation will create extensive socioeconomic opportunities by promoting the reduction of energy-related CO2 emissions and the decarbonization of societies. The speed with which this transformation is adopted and implemented as policy will determine the degree to which environmental hazards are reduced. This transformation is critical to achieving internationally agreed targets that require energy-related CO2 emissions to be reduced to 70% below current levels by 2050 (IRENA, 2020). Measures to reduce CO2 emissions should be designed specifically to increase the use of renewable energy and energy efficiency, because more than 90% of these measures rely on these two areas (Alola et al., 2019). The marked slowdown in the rate of improvement in energy intensity suggests that energy efficiency and renewable energy need to be significantly increased. The rate of energy development is expected to increase by 3.2% per year. This represents an increase of almost three times historical values (Saint Akadiri et al., 2019). To enable this transition, the transformation of the energy sector needs to be addressed not only from an environmental perspective but also from an energy policy and security perspective.

Energy security is a critical issue that directly impacts national security and causes major changes in foreign relations and world geopolitics. Energy security should be considered not only as protection against military aggression but also in a broader context of economic, food, energy, and environmental security. This broader framework includes an adequate response to the various risks of energy security and aims to protect the stability of the country and its citizens (Mohsin et al., 2019b). Modern energy security assessment has evolved to include not only national security considerations but also environmental and social concerns. The transformation of the energy sector emphasizes that energy policies should aim to balance energy security, economic development, and environmental sustainability.

In view of the preceding, the energy trilemma involves three conflicting challenges of balancing security, sustainability, and affordability (WEC, 2018). A country's ability to meet its energy needs in a reliable and sustainable manner depends on ensuring energy equity and environmental quality. In this regard, the reliability and resilience of the energy system are critical for ensuring energy security. Within the framework of the energy trilemma concept, the imperative to balance energy security, social impacts, and environmental considerations will provide a guide for effective planning and implementation of energy policies. In this way, the transformation of the energy sector towards a sustainable future will be supported and promoted both nationally and internationally. However, the management of the energy trilemma must be designed and implemented in accordance with the specific circumstances of each country. Therefore, in this study, OECD countries are analyzed in the context of the energy trilemma index and countries with similar characteristics are clustered. The results of the study are critical for energy policy makers and researchers to develop specific strategies to make each country's energy system more secure, sustainable, and accessible. These strategies will help reduce the vulnerabilities of national energy systems and address all three aspects of the energy trilemma in a balanced manner. In this way, countries that can effectively manage the energy trilemma will set an example for their citizens and the global community by taking important steps towards a sustainable energy future. This will not only accelerate the transformation of the energy sector but will also be of strategic importance for energy policy and security.

This study can contribute to the literature in at least 5 ways: (1) To our knowledge, no study has classified OECD countries based on the energy trilemma index. (2) Methodological diversity is provided using various statistical techniques such as the K-Means algorithm, the Elbow method, and the silhouette coefficient. (3) Using the most recent data up to 2022 provides a real-time and up-to-date perspective on the energy situation in OECD countries. (4) Visualization of research findings facilitates comprehension of the results and is an example of innovative and effective ways of presenting such analyses. (5) The study's findings provide a valuable resource for policy-making processes. These contributions can guide further research and serve as a reference point to better comprehend and manage the energy trilemma.

## 2. LITERATURE

Measuring energy performance and sustainability is crucial for the development of energy policies. In this regard, the Energy Trilemma Index (ETI) released by the World Energy Council (WEC) is a tool that comparatively assesses countries' performance in the dimensions of energy security, energy equity, and environmental sustainability (WEC, 2018). The assessment and analysis of the ETI is critical for measuring the effectiveness of energy policies and determining future strategies (Zou & Shen, 2023; Ponomarenko et al., 2022). Fu et al. (2022) emphasize that ETI is an important decision support tool that assesses energy security, equity, and environmental sustainability criteria. Moreover, Šprajc et al. (2019) state that ETI values are particularly high in the European Union and North America. This indicates that energy policies in these regions are managed in a more balanced and effective manner in terms of these three criteria. Lin et al. (2020) explain how the ETI is used to measure country-level energy performance and assess the sustainability development of the energy sector.

There has also been some criticism of the reliability and inclusiveness of the ETI. Šprajc et al. (2019) argue that the lack of methodological developments and differences in preferences across countries call into question the reliability of the ETI. Asbahi et al. (2019) discuss how different weighting methods can affect the ETI results and argue for the use of Stochastic Multi-Criteria Acceptability Analysis. In contrast, Zou & Shen (2023) argue that ETI can help policymakers balance the energy trilemma and contribute to policy recommendations by providing important insights in achieving this balance. Similarly, Kartsonakis et al. (2021) argue that the ETI framework can provide valuable policy recommendations for sustainable development. Chi et al. (2023) and Khan et al. (2021) explore the relationship between ETI and economic progress and emphasize the positive effects of energy transition processes on economic progress, in particular the necessity of transitioning to low-carbon energy sources. Ponomarenko et al. (2022) examine how ETI assesses energy sustainability, and in particular the potential of solar energy to increase energy security, affordability, and environmental sustainability.

In the literature, various methodologies other than the energy trilemma index have been used to analyze the energy performance of countries. For example, studies such as Song et al. (2017) evaluated national energy performance using methods such as Stochastic multi-criteria acceptability analysis. Moreover, statistical techniques such as Principal Component Analysis (PCA) have been used to assess energy performance and have found wide application (see, for example, Merton, 1973; Chamberlain & Arbitrage, 1982; Bei & Cheng, 2013; Zhang et al., 2015; Lever, 2016). Data Envelopment Analysis (DEA) is another important tool known for its applications in energy and environmental modelling. DEA is recognized as a powerful method for measuring environmental efficiency at the macroeconomic level and has been used in several studies to assess performance in the energy sector (Lovell et al., 1995; Seiford & Zhu, 2002). Zhou et al. (2008) provided a comprehensive literature review on how DEA can be used to measure energy efficiency and environmental performance. These techniques have made significant contributions in extracting important information from multidimensional data sets and evaluating energy policies (Topcu & Payne, 2017; Yoon & Klasen, 2018). Zhou & Ang (2008) developed a new DEA model that takes into account the undesirable outputs of economic activities. This model provides a broader perspective to energy efficiency analyses by taking environmental factors into account. In addition, Guo et al. (2017) put forward a dynamic DEA model that takes into account changes over time and analyzes managerial efficiency in OECD countries and China with time series data. Another study was conducted by Wang et al. (2017). In this study, an extended nonparametric frontier method that takes into account sectoral differences in the economy was developed, making it possible to analyze energy efficiency performance on a sectoral basis in detail. Focusing on the use of DEA techniques in energy efficiency assessments, Mardani et al. (2017) and Yu & He (2020) published comprehensive literature reviews on this topic. These reviews reveal how effective DEA is in energy efficiency analysis. Furthermore, Zhou et al. (2012) conducted a study that evaluated energy efficiency performance with a parametric approach, developed an index for energy efficiency using the Shephard distance function, and tested this index with stochastic frontier analysis. This study shows the potential of parametric methods in energy efficiency assessments. The index decomposition and spatio-temporal decomposition techniques developed by Ang et al. (2015) stand out as important tools for understanding how energy performance varies over time and space. Finally, Fu et al. (2021) developed a group decision-making methodology based on the Energy Architecture Performance Index. This methodology offers the possibility to analyze how energy policies are perceived and accepted by various stakeholders. Efforts to measure energy security and performance are also supported by international energy institutions and organizations. Organizations such as IEA, EECA, NRC, OEERE, and ODYSSEE have developed measurement and monitoring systems in the field of energy security and performance, providing policy makers with tools to use in their decision-making processes (Zhou & Ang, 2008).

Along with studies analyzing the energy performance of countries, there are also studies that cluster countries based on various energy indicators. In this respect, studies on energy consumption cover a wide range of methodological approaches. It is worth noting that research on this topic uses a range of analytical techniques to better comprehend and categorize the energy consumption profiles of different countries. For instance, Csereklyei et al. (2017) investigated the relationship between energy components and national income by examining the evolution of energy components of European Union member states over time. Similarly, Hu et al. (2018) analyzed the global energy consumption structure with the k-means classification algorithm and revealed the potential to diversify and improve the energy consumption structure of 144 countries. Jalali Sepehr et al. (2019) focused on energy efficiency and argued that changes in energy-related indicators can be an important factor in classifying countries. Finally, Gupta (2023) uses a simplified clustering approach to classify countries on the basis of their renewable energy consumption. This approach provides findings in terms of sustainability practices by revealing differences in renewable energy consumption by income level of countries.

Although these studies have categorized countries based on various energy indicators, there is no study in the literature that clusters countries by energy trilemma index. In this regard, the current research aims to cluster OECD countries based on the energy trilemma index and its dimensions. This approach aims to fill this gap in the literature and contribute to the field.

## 3. DATA AND METHODOLOGY

### 3.1 Model specification and data

The energy trilemma index is a tool used to assess the performance of a country's energy system. In this framework, the World Energy Council (WEC) publishes the energy trilemma index (ETI) report annually to assess energy sustainability. This index assesses the performance of countries' energy systems in three key dimensions: energy security, energy equity, and environmental sustainability. The ETI was prepared in collaboration with Oliver Wyman, a global consulting firm, and Marsh & McLennan Companies (WEC, 2015). In the 2015 ETI analysis conducted by WEC, it was observed that energy security, energy equity, and environmental sustainability were assessed in an integrated manner. These three dimensions provide a comprehensive assessment of countries' energy systems, revealing their energy policies' effectiveness and potential for progress in achieving sustainability goals. The ETI addresses the security, equitable access, and environmental impacts of energy systems in a balanced manner, emphasizing that this balance is critical for sustainable development (Hunter et al., 2016; Rempel et al., 2016). The components of the energy trilemma index are shown in Table 1.

**Table 1:** Components of energy trilemma index

| DIMENSION | WEIGHT | VARIABLES |
|---|---|---|
| Energy security | 30% | Import independence, Diversity of electricity generation, Energy storage |
| Energy equity | 30% | Access to electricity, Electricity prices, Gasoline and diesel prices |
| Environmental sustainability | 30% | Final energy intensity, Low carbon electricity generation, CO2 emissions per capita |
| Country context | 10% | Macroeconomic stability, Effectiveness of government, Innovation capability |

The energy trilemma index has 3 dimensions. The energy security dimension weighs 30% in the index and assesses a country's ability to meet its energy demand reliably. In this context, it considers factors such as the reliability of existing energy infrastructure, security of supply, resilience to energy shocks, import independence, diversity of electricity generation, and energy storage. Therefore, energy security is critical for economic growth, national security, and the stability of energy systems. The second dimension of the ETI, energy access, assesses a country's ability to provide access to energy resources and make energy available to consumers at fair and affordable prices. This is because access to energy is a fundamental requirement for ensuring social welfare and economic development. Therefore, this dimension considers factors such as the extent and accessibility of energy infrastructure and financial accessibility. The third dimension of the ETI, environmental sustainability, assesses the environmental impacts of a country's energy production and consumption processes. This is because environmental sustainability is essential for combating climate change, protecting natural resources, and maintaining ecological balance. In this regard, environmental factors such as greenhouse gas emissions, air and water pollution, and consumption of natural resources are considered. In addition to these three main dimensions, country context is also included in the index calculation. Macroeconomic and governance conditions, the stability of the economy and government, the attractiveness of the country to investors, and innovation capacity are assessed under this dimension.

The energy trilemma index benefits from: (1) It is an essential tool for understanding the complex nature of energy systems and the intersections of energy policies. (2) It enables countries to make a holistic assessment of their energy policies and analyze the current state of their energy systems. (3) It allows countries to track the evolution of their energy performance over time by providing comparative analyses across countries regarding energy security, equity, and environmental sustainability. (4) It can guide strategic decision-making on issues such as diversifying energy sources, expanding energy access, and reducing environmental impacts. (5) It can guide the financial sector in planning and directing investments in the energy sector. It is essential for investments in renewable energy and energy efficiency projects. (6) It contributes to developing international cooperation and policy dialogues. Thus, it promotes energy policy coordination and cooperation at the international level by identifying common challenges and opportunities facing the energy systems of different countries.

### 3.2 Model specification and data

Cluster analysis is an essential statistical method that emerged in the 1930s and is used in various disciplines. The k-means clustering method is a frequently preferred method that aims to divide data sets into meaningful subgroups and provides convenience in practical application (Aldenderfer & Blashfield, 1984; Xu & Wunsch, 2009).

The k-means clustering method has found a wide place in the literature as a clustering method that aims to group data sets into clusters of a determined number. This method seeks to maximize the intra-cluster similarities and minimize the inter-cluster similarities by dividing the data into groups according to their characteristics. The basis of the method is based on k randomly selected center points, and the data points are clustered closest to these centers, and this process continues until the centers are fixed (Evans et al., 2005; Amasyalı & Ersoy, 2008). K-means is a partitioning algorithm that belongs to the unsupervised learning category and can assign each data point to only one cluster because it has a sharp clustering mechanism. The steps of the K-means clustering method are summarized as follows (Steinbach et al., 2000; Karypis et al., 2000; Risheh et al., 2022):

**Step 1:** Determine the number of clusters (k) to be grouped: The number of objects selected k represents the cluster centers. The sample midpoint is calculated as in Equation 1 (Gersho & Gray, 1991).

$$M_k = \frac{1}{n_k}\sum_{i=1}^{n_k} x_{ik} \quad (1)$$

**Step 2:** Calculating within-cluster variation: The Squared Error Formula is calculated as in Equation 2 (Linde et al., 1980).

$$e_i^2 = \sum_{i=1}^{n_k}(x_{ik} - M_k)^2 \quad (2)$$

For the space of all clusters containing cluster K, the squared error is the sum of the changes within the cluster. The square-error value is then calculated as in Equation 3:

$$E_k^2 = \sum_{k=1}^{K} e_k^2 \quad (3)$$

**Step 3:** Assigning the data in the data set to the cluster with the cluster center closest to itself

**Step 4:** Once the clusters are identified (Step 3 is completed), calculate the cluster means and assign the obtained means as new starting centers

**Step 5:** Repeating Step 3 and Step 4 until the data in the clusters are fixed.

The success of the k-means algorithm depends on critical parameters such as the correct determination of the k value and the choice of starting points (Mohamed & Çelik, 2022). Incorrect determination of this value may negatively affect the reliability of the analysis results and may lead to inaccurate results. In the literature, various methods have been proposed to determine the optimal number of clusters, such as the Elbow method, Calinski & Harabasz's (1974) index, Davies & Bouldin's (1979) index, Krzanowski & Lai's (1988) index, and Silhouette's (1987) index. In this study, the elbow method and Silhouette (1987) index were used to determine the optimum number of clusters, and the results obtained were compared.

### 3.3 Elbow method

Determining the optimal number of clusters is critical in data analysis, especially clustering analysis. This is necessary to better comprehend the structure of the dataset and to reveal natural clustering between the data. One of the most frequently used methods in clustering analysis is the Elbow method. This method provides a heuristic approach to determining the optimal k clusters in a data set. The elbow method is an effective method for choosing the value of k in k-means analysis (Syakur et al., 2018; Taşçı & Onan, 2016).

The basis of the Elbow method is the calculation of the sum of the squares of the distances of each data point to the centers of the clusters. When this sum is calculated for different values of k, an "elbow point" emerges where the rate of decrease of this sum at a given value of k drops significantly. This point represents the optimal number of clusters from which the dataset can decompose naturally (Ketchen & Shook, 1996; Bholowalia & Kumar, 2014). The implementation of the Elbow method is as follows step by step (Taşçı & Onan, 2016; Syakur et al., 2018; Coşkun et al., 2021):

**Step 1:** Calculating the sum of squares within the cluster (WCSS) value for each k starting with k=1

**Step 2:** Recalculating the WCSS value by increasing k

**Step 3:** Identify the point at which the reduction of the WCSS decreases significantly. (This point at which the reduction decreases significantly is the elbow point at which the optimal number k is determined.)

**Step 4:** Determine the value of k at which the WCSS decreases less, and the modeling would not be improved by adding a cluster.

The Elbow method provides an idea of how best to cluster the dataset, even when visualization is limited if the dataset is multi-dimensional. This method aims to maximize the internal integrity and separation of clusters by evaluating the distance between the two most distant observations in the dataset.

### 3.4 Silhouette index

Another method frequently used in cluster analysis is the Silhouette index. The Silhouette Index is a value that measures how healthy clusters are defined, that is, how close the data within clusters are to each other and how far the data between clusters are from each other. The Silhouette index also called the shadow statistic, was developed by Rousseeuw in 1987. This coefficient shows which objects fit better into clusters. For any *data i*, the Silhouette value *s(i)* is calculated as in Equation 4 (Rousseeuw, 1987).

$$s(i) = \frac{b(i) - a(i)}{max\{a(i), b(i)\}} \tag{4}$$

The Silhouette value is calculated for each data point. This calculation is performed after clustering for a given value of k. The Silhouette value of a data point *i* is based on a(i) and b(i). Where a(i) is the average distance of data point i from all other data points in its cluster, and b(i) is the smallest value of the average distance of data point i from all data points in other clusters. The Silhouette value calculated using these two values indicates a data point belonging to its cluster and its separation from different clusters (Rousseeuw, 1987; Şenol & Karacan, 2020: 349). Silhouette values take values between -1 and 1. When the Silhouette value of a data point is close to 1, it indicates that the data point fits well into its cluster and is well separated from other clusters. When the Silhouette value is close to 0, it suggests it is unclear which cluster the data point belongs to. A negative Silhouette value indicates that the data point is probably assigned to the wrong cluster (Yılancı, 2010; Sangaiah et al., 2023; Verma et al., 2023).

## 4. RESULTS AND DISCUSSION

The Pair Plot analysis of OECD countries with the energy trilemma index allows for a detailed examination of the distribution of each variable and the relationships between four key variables: energy security, environmental sustainability, energy justice, and the energy trilemma index: energy security, energy equity, environmental sustainability, and the energy trilemma index. The visualization in Figure 2 illustrates the complex nature of energy policies and the interactions between the different dimensions of these policies.

**Figure 1:** Pair Plot Analysis Results

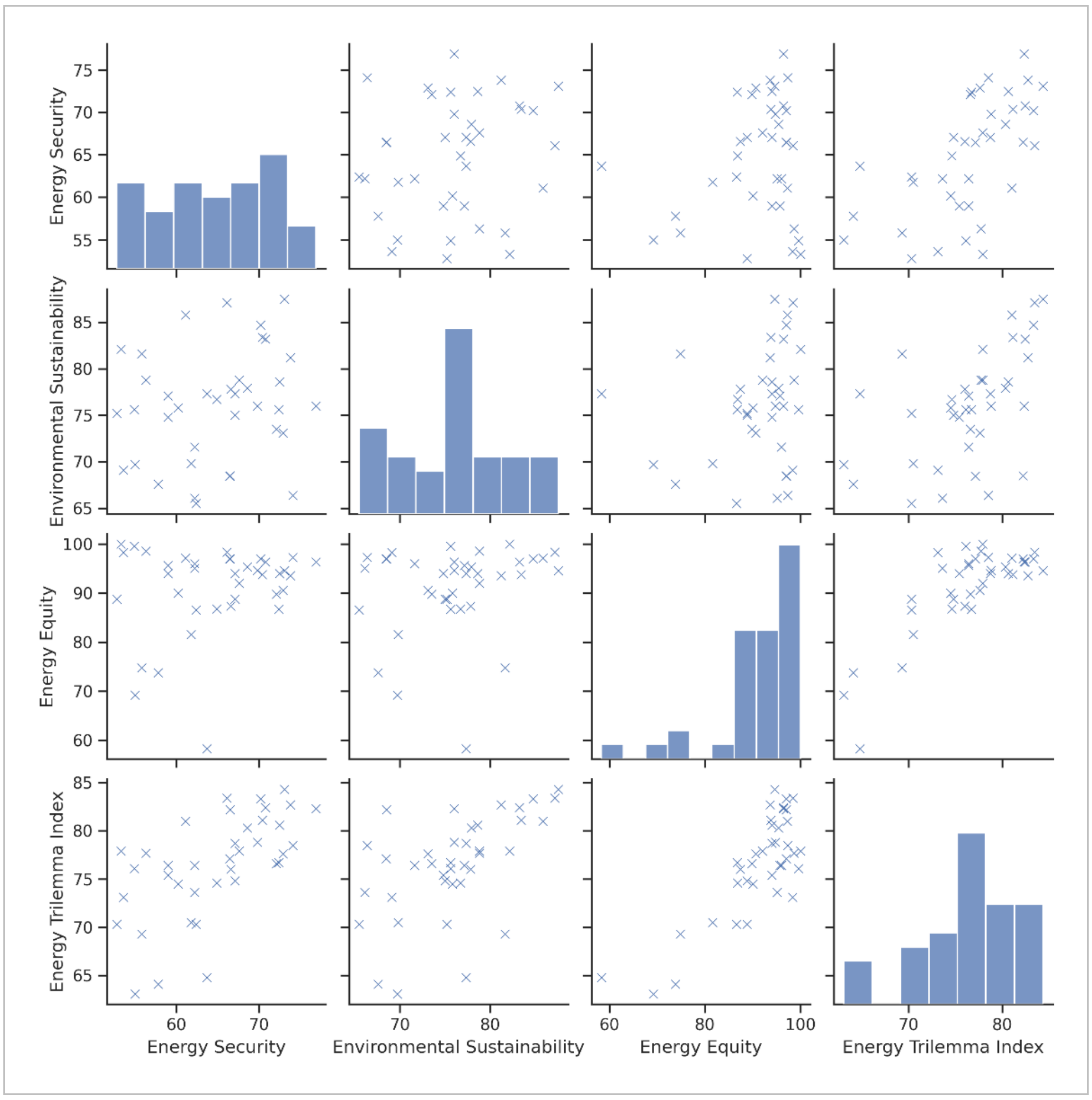

The positive relationship between energy security and environmental sustainability suggests that these two dimensions can support each other. In particular, investments in renewable energy sources can both increase energy security (by reducing external dependence) and improve environmental sustainability (by reducing carbon emissions and other environmental damages). The observation of this relationship highlights the importance of integrating environmental and security objectives in energy policies.

Energy equity is related to access to and equitable distribution of energy. The results suggest that the relationships between energy justice and other variables appear to be more complex. Countries with high energy equity generally have high values in the energy trilemma index. However, these variables are not directly related to environmental sustainability and energy security. This suggests that energy access and equitable distribution of energy are important factors affecting overall energy performance but do not by themselves establish strong relationships with other variables. Improving energy equity is important to increase energy access, especially for low-income households, and to distribute energy costs fairly.

The energy trilemma index assesses the energy performance of countries as a whole by combining three key dimensions: energy security, energy equity, and environmental sustainability. The fact that this index is positively correlated with the other three dimensions reveals the importance of balancing these three dimensions in energy policies. Having a high energy trilemma index value indicates that a country adopts a comprehensive and balanced approach to energy.

**Figure 2:** Elbow Method Results

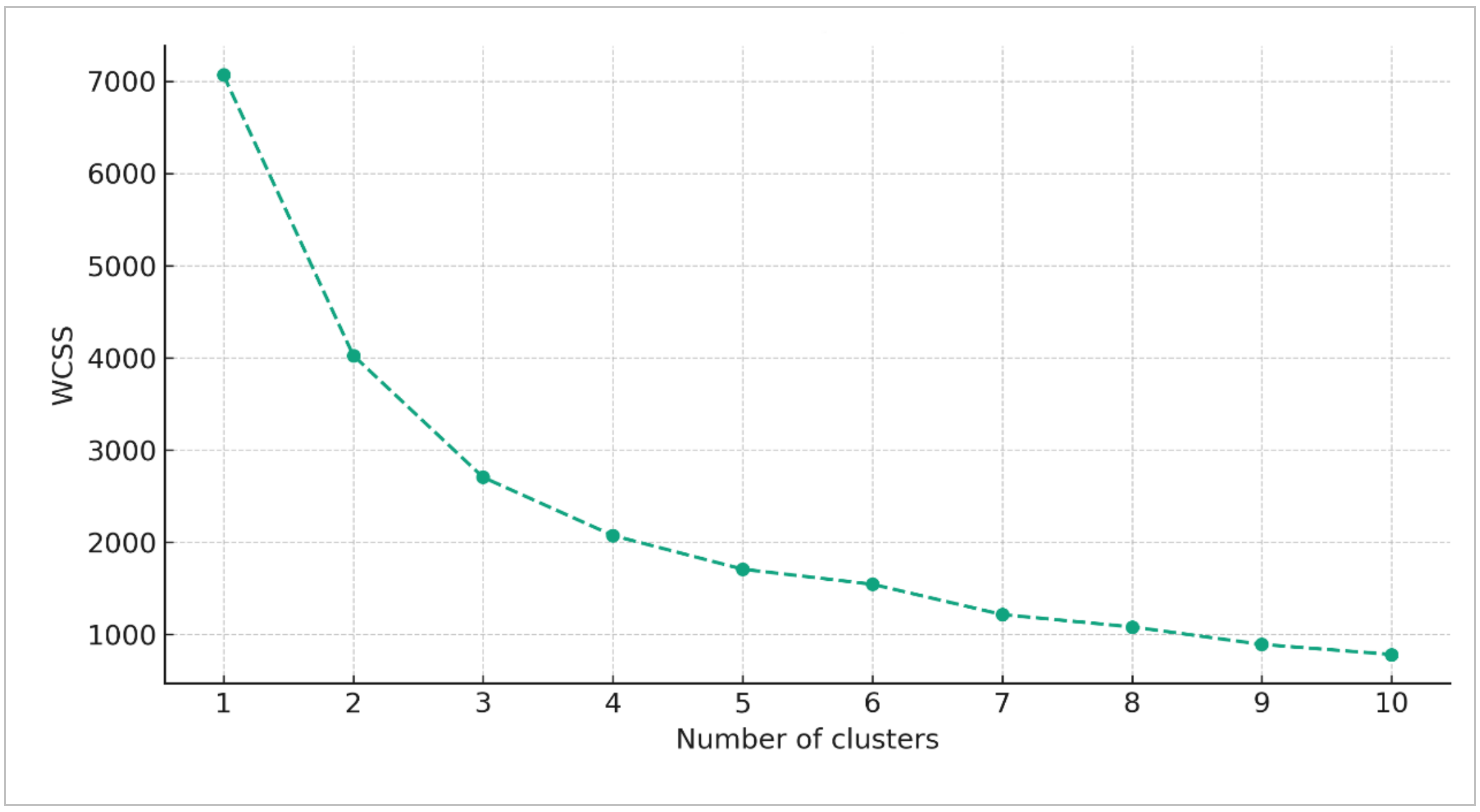

Figure 2 shows the result of the Elbow method. The results demonstrate that the elbow point is between clusters 3 and 4. Therefore, the optimum number of clusters can be chosen as 3 or 4. In this case, Silhouette Index was used to determine which cluster number provides better clustering performance. The silhouette coefficient for 3 clusters is approximately 0.354, and for 4 clusters it is approximately 0.312. In this regard, it is concluded that 3 clusters offer better clustering performance than 4 clusters. Therefore, the study continued with 3 clusters.

**Figure 3:** Dendrograms of Indices

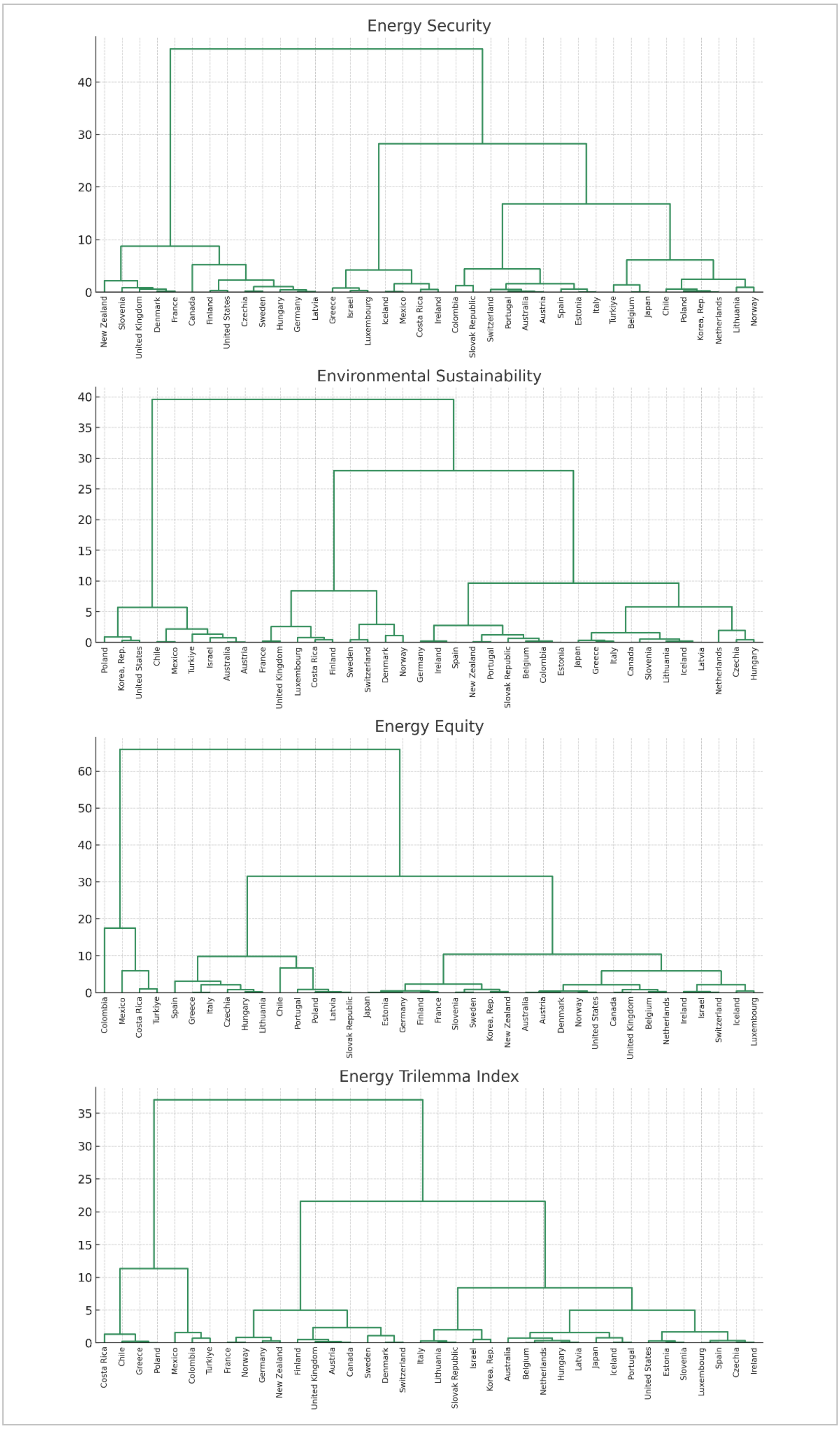

Figure 3 shows the dendrogram outputs of the indicators. In the dendrogram, the height of the joins (side vertical lines) represents the joins between clusters. The height of each union indicates the difference between clusters, with longer lines representing more significant differences. When a high line joins two clusters, they are less similar to each other than clusters joined by shorter lines. In this sense, the energy security dendrogram groups countries based on criteria such as access to energy resources, continuity of energy supply, and reliability of energy infrastructure. For example, Norway and the Netherlands are on the far right, indicating that they have similar and possibly high scores on energy security. Meanwhile, countries like South Korea and Mexico, which are in the same cluster in the graph, have similar characteristics in terms of energy security. However, similar to Norway and the Netherlands, they have a low similarity value.

The environmental sustainability dendrogram categorizes countries regarding energy policies and technologies that seek to minimize environmental impacts. Japan and Slovenia have similar approaches in this area. However, the similarities between countries grouped with a lower similitude, such as Hungary and Lithuania, are less pronounced compared to these two countries. The energy equity dendrogram shows countries in clusters according to factors such as energy cost and public access. Luxembourg and Switzerland are categorized with a very high similarity in terms of energy equity. This indicates equity and fairness in energy access. In addition, Colombia and Costa Rica have a lower similarity, indicating a different level of energy equity. Finally, the energy trilemma index dendrogram expresses an overall energy performance encompassing energy security, energy equity, and environmental sustainability. Here, countries such as Switzerland and Luxembourg are grouped with a high similitude, indicating a comprehensive success in their energy policies. In contrast, countries at the other end of the graph, such as the Czech Republic and Poland, which have a lower similitude in terms of the energy trilemma, indicate a lower level of alignment on this index. In each dendrogram, the within-group and between-country similitudes are important. For example, in the energy security dendrogram, the level of symmetry between Germany and other countries in its cluster (e.g., the United Kingdom and Canada) indicates that these countries have similar strategies for energy security. In contrast, its distance from other groups suggests that they adopt different approaches to energy security.

**Figure 4:** Energy Security Indices of Countries

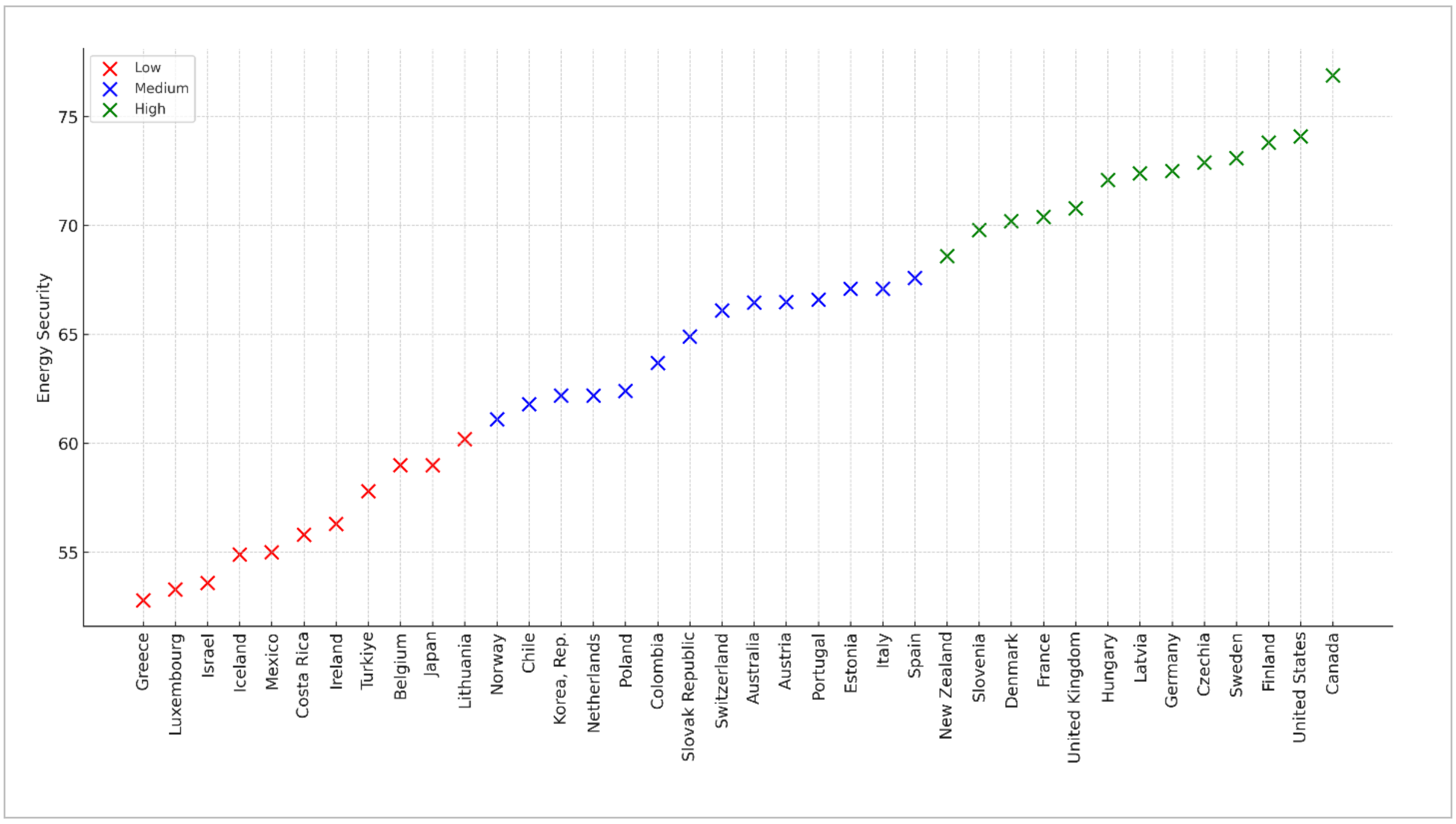

Figure 4 presents the clustering results of the energy security index. The high energy security index cluster includes countries with the highest energy security scores. Countries in this cluster outperform other countries in terms of factors such as diversity of energy sources, reliability of energy supply, and resilience of energy systems. These countries have generally made significant progress in reducing external dependence on energy supply and integrating renewable energy sources. For example, countries such as Canada, the US, Finland, and Sweden have high energy security scores, indicating that they have broad access to energy resources and strong infrastructure. These results are similar to those of Zhilkina (2019) and Fuentes et al. (2020). This can be attributed to their investments in renewable energy sources and focus on sustainability in their energy policies.

Countries in the medium energy security index cluster are between the first and third clusters in terms of energy security performance. While these countries have made some efforts to improve their energy security, further improvements and upgrades are needed. For example, countries such as Australia, Portugal, and the Slovak Republic continue to develop their energy infrastructure. However, a stable energy supply in these countries can only be sustained through dependence on energy imports. Therefore, while some of these countries face specific energy security challenges, they also have significant potential. For countries in this category, diversifying energy policies and developing strategies to enhance energy security should be a priority.

The low energy security index cluster includes countries with the lowest energy security scores. For example, countries such as Greece, Luxembourg, Israel, and Türkiye are included in this cluster. Countries in this cluster face serious challenges regarding the continuity and reliability of energy supply. Inadequate energy infrastructures, high energy import dependency, difficulties in transitioning to renewable energy sources, limited diversity of energy sources, infrastructure problems, or political and economic instability are among the main problems faced by these countries. Countries in this group need to undertake comprehensive reforms and diversify their energy sources to enhance energy security. These results are similar to Khatod et al. (2022) and Justus & Mannish (2023).

**Figure 5:** Energy Equity Indices of Countries

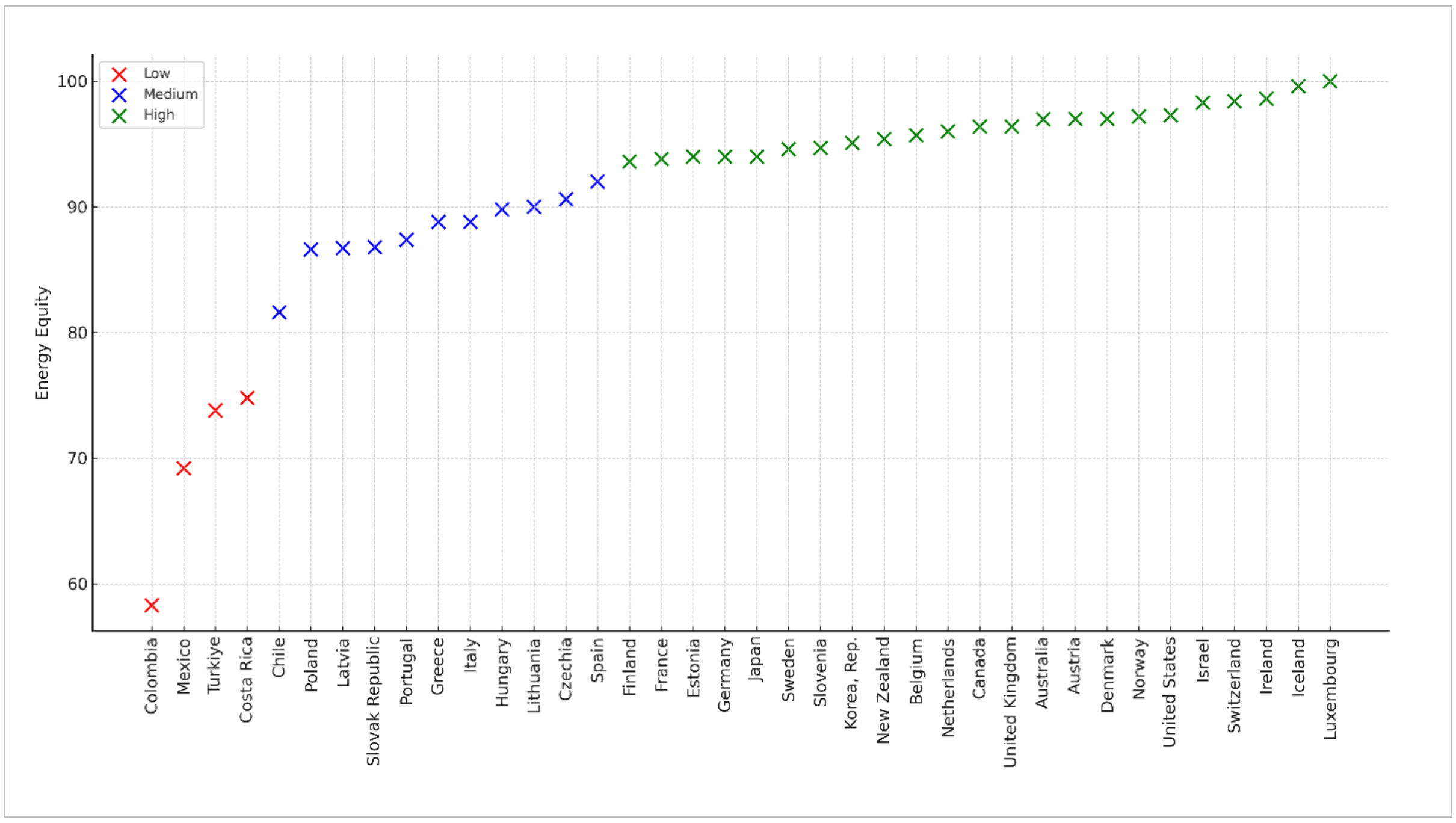

Figure 5 presents the clustering results of the energy equity index. Countries in the high energy equity index cluster are those where people have easy access to energy resources, energy costs are reasonable, and energy policies cover all levels of society. Countries in this category include high-income countries such as Luxembourg, Ireland, and Switzerland. These countries are able to provide energy resources in a fair and equitable manner to almost all their citizens. This suggests that higher income levels generally lead to a more equitable distribution of energy access.

Countries in the medium energy equity index cluster are those with a balanced performance in terms of access and cost of energy resources. While this indicates a more balanced distribution of energy resources and wider access to energy, it also suggests that there may still be significant differences in energy access among individuals in these countries. Countries with a medium energy equity index include Poland, the Slovak Republic, and Italy. Poland is struggling with its over-reliance on coal for energy production and its environmental impacts. This is a limiting factor for energy equity. The Slovak Republic has a medium index due to economic transformations and low competition in the energy market. In Italy, old and inadequate energy infrastructure and a slow transition to renewable energy point to a moderate level of energy equity. While these countries face challenges in energy access and distribution, these are not as extreme as in low-index countries.

Countries in the low energy equity index cluster are those where energy costs are a heavy cost burden for the population, where there are severe restrictions on access to energy, or where certain segments of society are excluded from energy sources. Countries in this cluster include Colombia, Mexico, and Türkiye. In Colombia, inadequate energy infrastructure in rural areas and inequitable distribution of energy resources negatively affect the energy equity index. In Mexico, corruption, underinvestment in the energy sector, and inefficient distribution systems contribute to increasing inequalities in energy access. Moreover, in Türkiye, energy price volatility and high import dependency pose energy equity challenges. The main problems in these countries can be summarized as infrastructure deficiencies, high energy costs, and energy market instability.

**Figure 6:** Environmental Sustainability Indices of Countries

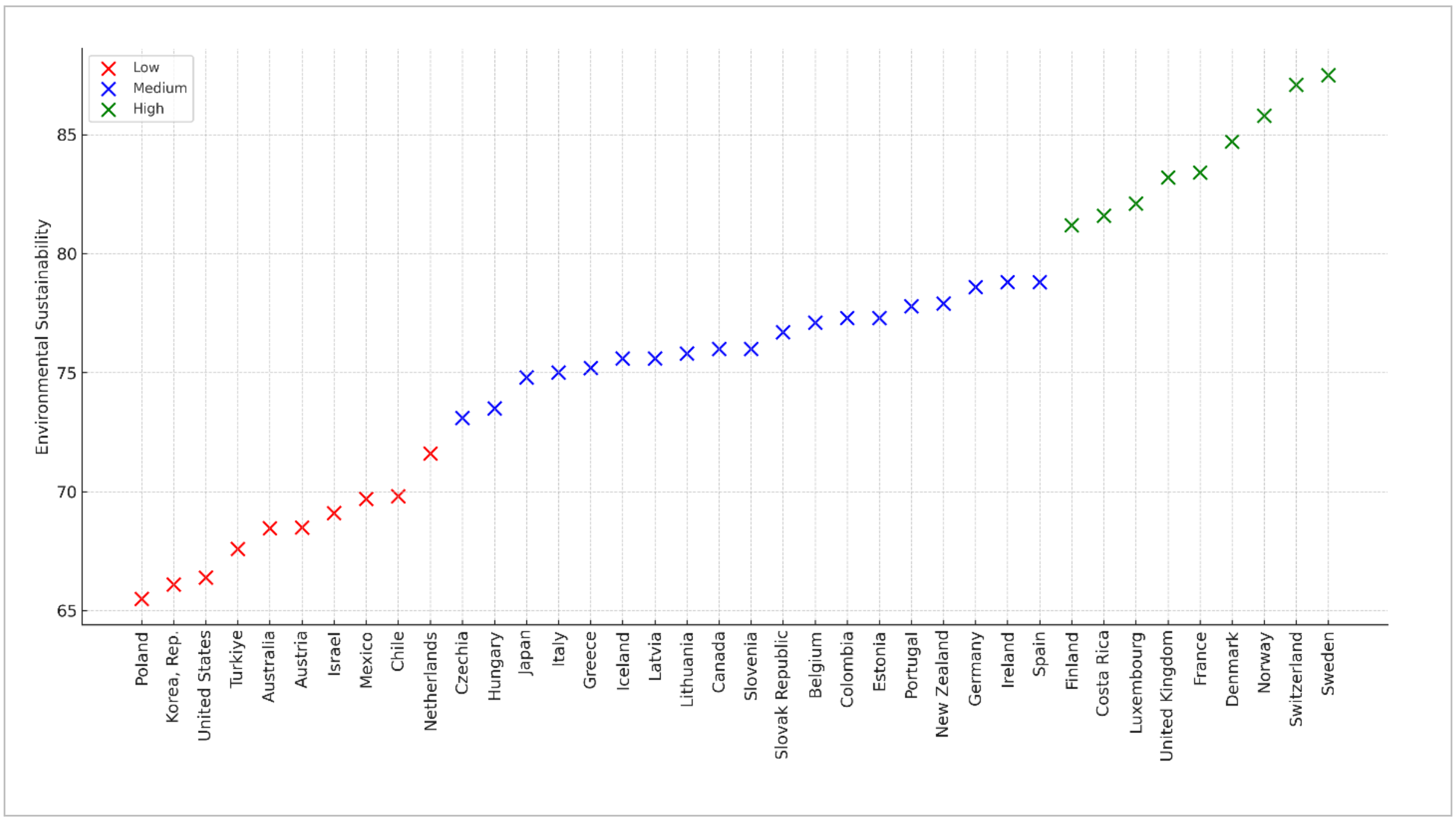

Figure 6 shows the clustering of OECD countries based on the environmental sustainability index. The high environmental sustainability index cluster includes countries with the highest scores in terms of environmental sustainability. These countries stand out with low carbon emissions and high renewable energy utilization rates. Countries in this cluster have successfully implemented policies that prioritize environmental sustainability, thereby significantly reducing their negative impact on the environment. For example, countries such as Sweden, Switzerland, and Norway are in this group. Their investments in renewable energy sources reflect their commitment to energy efficiency and the strength of their environmental protection laws (Garrido et al., 2020; Sueyoshi et al., 2022). As a result, countries in the high environmental sustainability cluster are exemplary in environmental sustainability and serve as models for other countries.

The medium environmental sustainability index cluster includes countries that consider environmental factors but do not achieve the highest scores. Countries such as Canada, Slovakia, and Belgium are in this cluster. Countries in this cluster have generally made significant strides in reducing carbon emissions and integrating more renewable energy sources. However, they have not fully transitioned to improvements in energy efficiency and sustainable energy sources.

The low environmental sustainability index cluster includes countries with the lowest scores in terms of environmental sustainability. This cluster includes countries that are often over-dependent on fossil fuels and underinvest in renewable energy sources. High carbon emissions and low levels of energy efficiency are the main reasons for their low environmental sustainability performance. This creates significant barriers to both combating global climate change and mitigating local environmental problems. For example, countries such as Poland, the Republic of Korea and Australia, Israel, and Türkiye are in this group. These countries are characterized by high dependence on fossil fuels, underinvestment in renewable energy, or weak environmental protection laws (see, for example, Kumar, 2020; Kumar et al., 2020). Improving environmental sustainability requires a global effort and requires countries to cooperate within and among themselves. In this regard, national and international policies, innovative solutions, and technological progress are crucial to enhancing environmental sustainability. Every step in this direction will make a significant contribution to building a more livable and sustainable world for present and future generations.

**Figure 7:** Energy Trilemma Indices of Countries

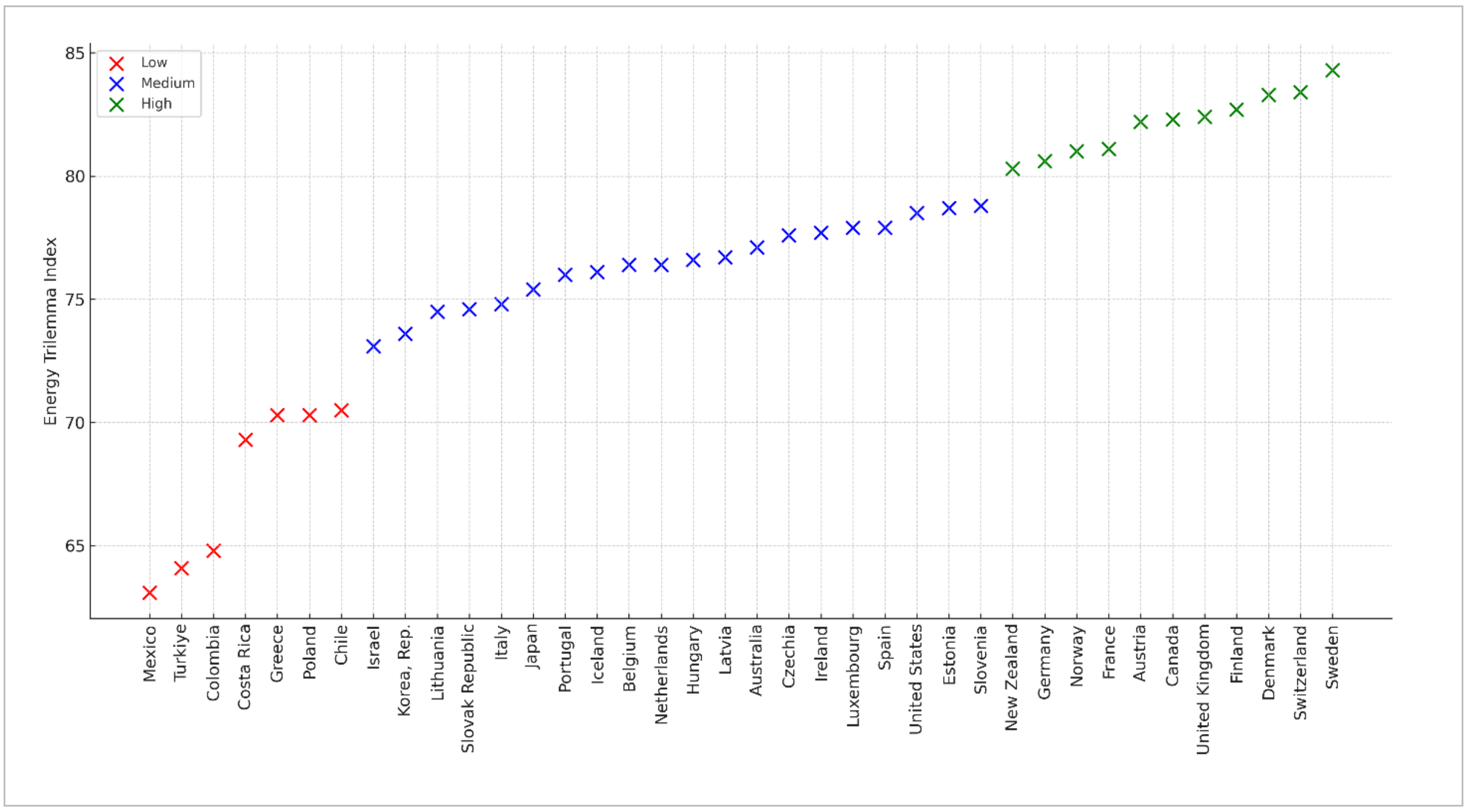

Figure 7 shows the clustering of OECD countries in terms of energy trilemma index. The cluster of high energy trilemma index includes countries that successfully manage the energy trilemma. Countries such as Sweden, Switzerland, and Denmark stand out in this cluster. With superior performance in the areas of energy security, environmental sustainability, and energy equity, these countries are recognized as global leaders in sustainable energy systems. Their energy policies focus on the efficient use of renewable energy sources, increasing energy efficiency, and reducing carbon emissions. They also implement comprehensive policies to expand access to energy and keep energy costs at reasonable levels. Countries in the high cluster promote innovation in the energy sector and emphasize international cooperation to achieve sustainable development goals. Countries in this cluster are role models for other countries in terms of energy policies and implementation.

Countries in the medium energy trilemma index cluster are those that achieve good results in one or two of the three pillars but do not perform at the top level in all three areas. For example, countries such as Italy, Japan, and the United States are in this cluster. Countries in this cluster have made significant progress in ensuring energy security, reducing environmental impacts, and ensuring equitable access to energy (see Figure 4, Figure 5, and Figure 6). However, these countries still face various challenges. For example, additional investments and policy changes may be required in areas such as integrating renewable energy sources, modernizing energy infrastructure, or improving energy efficiency. Countries belonging to this cluster should develop strategies to move to the next cluster by further optimizing their energy systems.

Countries in the low energy trilemma index cluster face serious challenges in terms of energy security, environmental sustainability, and/or energy equity. These challenges can take the form of interruptions in energy supply, high energy costs, or poor management of environmental impacts. For example, a country with low energy security may not have sufficient and reliable energy supply, which can negatively impact economic stability and development. Countries with low environmental sustainability scores fail to tackle environmental challenges such as carbon emissions and air and water pollution. Countries with low energy equity have economic or geographical barriers to energy access, deepening social injustice. Countries in this cluster need to overcome these challenges by reviewing their energy policies.

Successful management of the energy trilemma is vital for achieving the goals of sustainable development and environmental protection. Clustering each country based on the energy trilemma index provides guidance for assessing the effectiveness of energy policies and strategies and for future improvements. Therefore, in Figure 8, all index values and clusters for OECD countries are presented comparatively and interpreted in detail.

**Figure 8:** All Indices of OECD Countries

| | Energy Security | | | Energy Equity | | | Environmental Sustainability | | | Energy Trilemma | | |
|---|---|---|---|---|---|---|---|---|---|---|---|---|
| **Country** | **Index** | **Rank** | **Cluster** | **Index** | **Rank** | **Cluster** | **Index** | **Rank** | **Cluster** | **Index** | **Rank** | **Cluster** |
| Australia | 66.47 | 19 | Medium | 96.99 | 10 | High | 68.47 | 34 | Low | 77.1 | 19 | Medium |
| Austria | 66.5 | 18 | Medium | 97 | 9 | High | 68.5 | 33 | Low | 82.2 | 7 | High |
| Belgium | 59 | 29 | Low | 95.7 | 14 | High | 77.1 | 17 | Medium | 76.4 | 22 | Medium |
| Canada | 76.9 | 1 | High | 96.4 | 11 | High | 76 | 19 | Medium | 82.3 | 6 | High |
| Chile | 61.8 | 26 | Medium | 81.6 | 34 | Medium | 69.8 | 30 | Low | 70.5 | 32 | Low |
| Colombia | 63.7 | 22 | Medium | 58.3 | 38 | Low | 77.3 | 16 | Medium | 64.8 | 36 | Low |
| Costa Rica | 55.8 | 33 | Low | 74.8 | 35 | Low | 81.6 | 8 | High | 69.3 | 35 | Low |
| Czechia | 72.9 | 5 | High | 90.6 | 25 | Medium | 73.1 | 28 | Medium | 77.6 | 18 | Medium |
| Denmark | 70.2 | 11 | High | 97 | 8 | High | 84.7 | 4 | High | 83.3 | 3 | High |
| Estonia | 67.1 | 15 | Medium | 94 | 20 | High | 77.3 | 15 | Medium | 78.7 | 13 | Medium |
| Finland | 73.8 | 3 | High | 93.6 | 23 | High | 81.2 | 9 | High | 82.7 | 4 | High |
| France | 70.4 | 10 | High | 93.8 | 22 | High | 83.4 | 5 | High | 81.1 | 8 | High |
| Germany | 72.5 | 6 | High | 94 | 19 | High | 78.6 | 12 | Medium | 80.6 | 10 | High |
| Greece | 52.8 | 38 | Low | 88.8 | 29 | Medium | 75.2 | 24 | Medium | 70.3 | 33 | Low |
| Hungary | 72.1 | 8 | High | 89.8 | 27 | Medium | 73.5 | 27 | Medium | 76.6 | 21 | Medium |
| Iceland | 54.9 | 35 | Low | 99.6 | 2 | High | 75.6 | 22 | Medium | 76.1 | 24 | Medium |
| Ireland | 56.3 | 32 | Low | 98.6 | 3 | High | 78.8 | 10 | Medium | 77.7 | 17 | Medium |
| Israel | 53.6 | 36 | Low | 98.3 | 5 | High | 69.1 | 32 | Low | 73.1 | 31 | Medium |
| Italy | 67.1 | 16 | Medium | 88.8 | 28 | Medium | 75 | 25 | Medium | 74.8 | 27 | Medium |
| Japan | 59 | 30 | Low | 94 | 21 | High | 74.8 | 26 | Medium | 75.4 | 26 | Medium |
| Korea, Rep. | 62.2 | 24 | Medium | 95.1 | 16 | High | 66.1 | 37 | Low | 73.6 | 30 | Medium |
| Latvia | 72.4 | 7 | High | 86.7 | 32 | Medium | 75.6 | 23 | Medium | 76.7 | 20 | Medium |
| Lithuania | 60.2 | 28 | Low | 90 | 26 | Medium | 75.8 | 21 | Medium | 74.5 | 29 | Medium |
| Luxembourg | 53.3 | 37 | Low | 100 | 1 | High | 82.1 | 7 | High | 77.9 | 15 | Medium |
| Mexico | 55 | 34 | Low | 69.2 | 37 | Low | 69.7 | 31 | Low | 63.1 | 38 | Low |
| Netherlands | 62.2 | 25 | Medium | 96 | 13 | High | 71.6 | 29 | Low | 76.4 | 23 | Medium |
| New Zealand | 68.6 | 13 | High | 95.4 | 15 | High | 77.9 | 13 | Medium | 80.3 | 11 | High |
| Norway | 61.1 | 27 | Medium | 97.2 | 7 | High | 85.8 | 3 | High | 81 | 9 | High |
| Poland | 62.4 | 23 | Medium | 86.6 | 33 | Medium | 65.5 | 38 | Low | 70.3 | 34 | Low |
| Portugal | 66.6 | 17 | Medium | 87.4 | 30 | Medium | 77.8 | 14 | Medium | 76 | 25 | Medium |
| Slovak Republic | 64.9 | 21 | Medium | 86.8 | 31 | Medium | 76.7 | 18 | Medium | 74.6 | 28 | Medium |
| Slovenia | 69.8 | 12 | High | 94.7 | 17 | High | 76 | 20 | Medium | 78.8 | 12 | Medium |
| Spain | 67.6 | 14 | Medium | 92 | 24 | Medium | 78.8 | 11 | Medium | 77.9 | 16 | Medium |
| Sweden | 73.1 | 4 | High | 94.6 | 18 | High | 87.5 | 1 | High | 84.3 | 1 | High |
| Switzerland | 66.1 | 20 | Medium | 98.4 | 4 | High | 87.1 | 2 | High | 83.4 | 2 | High |
| Turkiye | 57.8 | 31 | Low | 73.8 | 36 | Low | 67.6 | 35 | Low | 64.1 | 37 | Low |
| United Kingdom | 70.8 | 9 | High | 96.4 | 12 | High | 83.2 | 6 | High | 82.4 | 5 | High |
| United States | 74.1 | 2 | High | 97.3 | 6 | High | 66.4 | 36 | Low | 78.5 | 14 | Medium |

Legend: Low (red) · Meddium (yellow) · High (green)

Figure 8 shows all indices calculated for OECD countries. Countries in the low energy trilemma index cluster stand out. There are seven countries in this cluster, including Mexico, Türkiye, Colombia, Costa Rica, Greece, Poland, and Chile. Among these countries, Greece has the lowest energy security (52.8) and Colombia the highest (63.7). These differences depend on factors such as access to energy resources, the balance of energy production and consumption, and dependence on energy imports. For example, Colombia's relatively high energy security depends on the country's ability to utilize its rich hydropower resources (see, for example, Bravo-López et al., 2022; Leguizamon-Perilla et al., 2023). The country with the highest environmental sustainability score in this cluster is Costa Rica (81.6). The country with the lowest environmental sustainability index is Poland (65.5). Costa Rica's high score in the environmental sustainability index is a result of the country's investments in renewable energy sources. Particularly noteworthy is the energy generated from hydroelectric, geothermal, and wind power. Poland's low score is due to the fact that it is still heavily dependent on coal-based energy production (see, for example, Woźniak & Pactwa, 2019; Dzikuć et al., 2021). In terms of energy equity, Colombia has the lowest energy equity score (58.3), while Greece has the highest score (88.8). The low energy equity score in Colombia points to inequalities in energy access and infrastructure problems. In Greece, the prevalence of energy subsidies and policies to regulate energy prices contributed to the high energy equity score.

Countries in the low energy trilemma index cluster perform differently in each component of the energy trilemma. This is a result of each country's unique energy profile, access to natural resources, and energy policies. The challenges these countries face in building a sustainable and equitable energy system are diverse. It is therefore crucial to adopt holistic and integrated policies to promote energy security, environmental sustainability, and energy equity. Policies that address the three components of the energy trilemma in a balanced way can help countries address their energy challenges in both the short and long term. These policies also play a critical role in combating climate change in line with international commitments such as the Paris Agreement. In particular, increasing investments in renewable energy generation sources, promoting energy efficiency, and modernizing energy infrastructure will increase environmental sustainability and energy security. In this respect, the transition to renewable energy is both a challenge and an opportunity, especially for countries that are heavily dependent on coal-based electricity generation. This transition will not only enhance environmental sustainability but also strengthen energy security by reducing dependence on energy imports. Moreover, ensuring transparency in energy pricing, effectively managing energy subsidies, and implementing policies to ensure equal access to energy can significantly improve energy equity. However, the socioeconomic impacts of this transition should also be taken into account, and social cohesion policies, such as the retraining of energy sector workers, should be developed. Furthermore, the efforts of these countries to stabilize the energy trilemma should be supported by international cooperation, knowledge, and technology transfer. Support from the international community can facilitate the achievement of these countries' sustainable energy goals, especially in terms of technology transfer and access to finance. Holistic approaches and multilateral cooperation are increasingly important for achieving global energy security, environmental sustainability, and energy equity. In this framework, the challenges and opportunities faced by these countries should be addressed not only at the national level, but also from a global perspective. This is because balanced progress on the three components of the energy trilemma is critical for these countries to achieve their development goals as well as to address global climate change.

## 5. CONCLUSIONS

The energy trilemma is recognized as one of the major challenges facing the energy sector worldwide. This complex approach, which aims to achieve a balance between energy security, energy equity, and environmental sustainability, is a fundamental framework that countries should take into account when shaping their energy policies. However, the management of the energy trilemma should be designed and implemented in accordance with the specific circumstances of each country. Therefore, this study analyzes OECD countries in the context of the energy trilemma index and clusters countries with similar characteristics. At the end of the study, solutions to the current and future challenges of the energy sector are explored.

The results show that countries with high energy security scores include Canada, the United States, Germany, Sweden, and Finland. These countries outperform others in terms of diversity of energy sources and reliability of energy systems. Countries with low energy security indices include Türkiye, Greece, Israel, and Mexico, which are characterized by energy supply disruptions and high energy import dependence. In terms of environmental sustainability, high-index countries such as Sweden, Switzerland, Norway, and Denmark stand out with their low carbon emissions and use of renewable energy. In contrast, countries with low sustainable environmental indices include Poland, Türkiye, the United States, Israel, and Chile. These countries are characterized by overdependence on fossil fuels and underinvestment in renewable energy sources. In terms of energy equity, inequalities in access to energy and challenges in the equitable distribution of energy stand out as major problems in Colombia, Mexico, Türkiye, and Costa Rica. This shows that ensuring fairness in energy policies should go beyond economic and technological factors to include social and cultural dimensions. However, countries with high scores on energy equity, such as Luxembourg, Switzerland, Australia, Austria, Ireland, and Iceland, lead by example by ensuring widespread access to energy and a fair distribution of energy costs. As a result, successfully managing the energy trilemma is an important imperative for countries in both the short and long term. Countries such as Sweden, Switzerland, Denmark, Finland, the United States, and the United Kingdom have successfully managed the energy trilemma by adopting a comprehensive and balanced approach to energy security, environmental sustainability, and energy equity. These countries have increased sustainability in the energy sector through investments in renewable energy sources, a focus on energy efficiency, and environmental protection policies. In contrast, countries such as Mexico, Colombia, Greece, Israel, Poland, and Türkiye face serious challenges due to their low energy Trilemma indices.

The findings of this study emphasize the importance of taking a holistic approach to the three dimensions of the energy trilemma. Therefore, the study developed a series of policy recommendations based on the results obtained: (1) Countries should increase their investments in line with their renewable energy potential. The first steps to be taken in this context include providing financial incentives for renewable energy projects and reducing bureaucratic obstacles. (2) It is important to increase energy efficiency. Therefore, it is recommended to raise energy efficiency standards in buildings, industry, and transportation sectors and increase incentives for energy efficiency projects. (3) Ensuring energy security is critical. Steps such as diversifying energy sources and building strategic energy reserves should be considered in this scope. (4) Energy assistance programs for low-income households should be developed to ensure energy equity. In addition, projects to improve energy infrastructure in rural areas should be prioritized. (5) Economic instruments such as carbon tax and investment in green technologies are recommended to reduce carbon emissions. (6) Technological innovation and R&D should be encouraged, investments in clean energy technologies should be supported, and cooperation between universities and the private sector should be increased. (7) International cooperation needs to be strengthened by increasing international cooperation on energy policies and technology transfer. (8) It is recommended to develop social cohesion policies, identify strategies for the transformation of the labor market in the energy sector, and increase employment opportunities in the renewable energy sector. (9) Public awareness and participation should be ensured for the success of policies. To this end, the public should be informed about energy conservation and environmental protection issues, and the active participation of local communities in the decision-making processes of energy projects should be encouraged. These recommendations can guide OECD countries in managing the energy trilemma and building a sustainable energy future.

Although this study has important findings, it also has some limitations. This study can be improved by addressing these limitations in future studies. First, this study uses the energy Trilemma Index calculated by the World Energy Council. The depth of the study can be increased by increasing the indicators to be used in future studies. Second, OECD countries were selected as the sample in this study. The sample of future studies can be selected from developed and developing countries, and the results obtained can be compared. Third, K-means and Elbow methods and the Silhouette index were used in this study. The study's results can be compared using different and hybrid techniques.